Ph.D Dissertation proposal

# A new compact, portable 2-LTD-brick *x*-pinch driver at the Idaho Accelerator Center: design, fabrication, testing and *x*-ray performance

Roman V. Shapovalov
Physics Department
Idaho State University

*Adviser:*      George Imel, Ph.D
*Co-adviser:*   Rick Spielman, Ph.D
*Committee:*    Mahbubul Khandaker, Ph.D
*Committee:*    Khalid Chouffani, Ph.D
*GFR:*          DeWayne Derryberry. Ph.D

October, 2014

# Contents

# ABSTRACT

There is a great interest in an *x*-pinch *x*-ray radiation source for use in a variety of applications where a bright, sub-nanosecond, and μm radiation source pulse is needed. Some examples include point-projection radiography in plasmas, imaging of soft biological objects, characterization of inertial-confinement-fusion capsule shells, and more. However, because almost all well-known, "standard" *x*-pinch radiation machines are based on a conventional Marx generator, they are large, expensive to build, operate, and are therefore not easily available for all laboratories where an *x*-pinch radiation source is required. Some small *x*-pinch generators exist, but they still lack portability.

Literature suggests that current rise rate, dI/dt, of 1 kA/ns or more is required for a "good" *x*-pinch radiation performance. For reasonable current rise times, this translates to a current requirement of 150 kA or more. We propose to construct a compact and portable *x*-pinch driver with *x*-ray radiation performance, comparable to standard *x*-pinch drivers. Such a new *x*-pinch driver was recently designed, fabricated and tested at the Idaho Accelerator Center. The generator is based on two slow LTD bricks combined into one solid unit, and can be described by a simple RLC circuit with four fast 140-nF, 100-kV capacitors that store up to 2.8 kJ. The test data reveals that when charged to 80 kV, the driver supplies 185-kA peak-current into a short Ni-wire load with 220-ns, 10-90%, rise time. The total internal inductance of our driver was measured to be about 60 nH. The revised driver model shows that when fully charged to 100 kV, the driver will supply 180-kA peak-current with 150-ns rise-time into the *x*-pinch load. The corresponding current rise rate is about 1.2 kA/ns.

To prove the driver *x*-pinch efficiency and to estimate the *x*-ray radiation performance, we could, for example, image an exploding wires, placed in a separate HV pulser, with our *x*-pinch *x*-ray radiation source. The study of exploding wires helps to understand the behavior of a warm dense matter, and our *x*-pinch driver can be part of the diagnostics needed for this study which is currently under progress at the IAC. Our driver contains no oil inside, is very compact and portable, and can be easily relocated to practically anywhere, which makes it an ideal backlighting diagnostic tool in many areas of plasma physics, biology, and industry where a bright, fast, and small *x*-pinch radiation source is required.

# 1. INTRODUCTION

The *x*-pinch, *x*-ray radiation source (in which the crossing and touching of two or more thin wires forms an "x", hence, *x* pinch) was first introduced by Russian physicists at Lebedev Physical Institute [1], and has since generated a great interest in many areas of plasma physics, biology, and industry. When a high, fast rising current is passing through *x*-pinch wires, a very fast, small and bright x-ray radiation source is produced. The width can vary from a few ps to tens of ns, the source size could be as small as a few μm, radiation power is on the order of a few hundred mJ, and the energy spectra ranges from soft to hard *x*-rays. All these radiation parameters are especially useful, when a high temporal and spatial resolution is desired to study many different, fast evolving, low opacity objects which makes the *x*-pinch, *x*-ray radiation source an unique diagnostic tool in many areas of researches. Some known *x*-pinch applications include, but not are limited to: point-projection radiography of exploding wires [2-8], backlighting of other *x*-pinch [9-13] or even *z*-pinch [14-16] systems, phase-contrast imaging of soft biological objects [6, 9, 17, 18], studying low density CH foams on an axis of wire array *z*-pinches [19], characterization of inertial-confinement-fusion capsule shells [20], or and even microlithography [21].

For a long time, a conventional pulsed-power device for driving an x-pinch load was a high-voltage Marx generator coupled with one or more transmission lines to compress an initially long (few μs) pulse from the generator's output into a short (a few hundred ns) pulse at the load. XP pulser [22, 23], for example, operating at Cornell University, consists of a 10-stage Marx generator, 4 coaxial intermediate storage capacitors, 3 coaxial pulse forming lines, one self-breaking gas switch, and 8 self-breaking water switches. It was designed and built about 20 years ago and can delivery about 450 kA with a 40-ns, 10-90%, rise time to a low inductance load. The Llampudken [24, 25] high-current pulser at Pontifical Catholic University in Chile consists of two Marx capacitor banks with a water transmission line coupled with each Marx generator, and it can supply about 400-kA peak current with a 260-ns rise time. Recently built at Tsinghua University in Beijing, China, PPG-1 [26, 27] is composed of a 1.2 MV Marx generator, pulse forming line, transmission line, switch, and a load section, and it can supply a 400-kA peak-current with a 100-ns pulse-width. Some other known pulsed-power installations used in *x*-pinch researches are LION [28], a 470-kA 80-ns pulser at Cornell; COBRA [29], a 1-MA, 100-ns generator at Cornel; Gamble II, a 1-MA, 100-ns pulser at the Naval Research Laboratory, in Washington DC; ZEBRA [30, 31], a 1.2-MA, 90-ns pulser at University of Nevada; MAGPIE [32, 33], a 1.8-MA 150-ns TeraWatt facility at Imperial College; S-300 [34], a 2.3-MA 150-ns high-current generator at Lebedev Physical Institute, in Moscow, Russia; and more. Examples of smaller *x*-pinch drivers, which can deliver a current up to 200 kA are a 80-kA 50-ns X-Pinch Pulser [35, 36] at the University of California, San Diego; 100-kA 60-ns table-top *x*-pinch [6] at Tsinghua University, China; GEPOPU [37], a 180-kA, 120-ns low inductance pulser at Pontifical Catholic University, Chile; and Light-II [38], a 200-kA pulsed-power generator at China Institute of Atomic Energy (CIAF), Beijing, China.

All such pulsed-power systems are able to deliver a large, fast rising current to a low inductance x-pinch load and have generated for decades a huge amount of *x*-pinch research around the world. But, in summary, all these *x*-pinch installations, are based on conventional Marx generators, water forming lines, and gas switches, all of which have many drawbacks. Some of them [22-34] are huge and bulky, and therefore are expensive to build and operate. Others [36-38] are more compact, but they are still require a conventional Marx generator filled with oil, water transmission lines, and gas system, which is often inconvenient. Large drawbacks of all such systems are that they still lack portability and are usually limited to one location to conduct experiments. Indeed, our initial design of *x*-pinch plasma-radiation-source generator [39] was based on an idea to non-destructively convert the Idaho State Induction System (ISIS) induction cell driver (ICD) to a low-impedance pulsed-power driver. Such a system would be composed of Marx generator, 5 pulse forming lines coupled to an impedance converter, and would supply about 200-kA peak-current with a 60-ns, 10-90%, rise time [39] into an *x*-pinch load. But, if constructed, it would be a large, fixed facility, located at the ISIS IAC site, which would be difficult to relocate to another place if needed.

Recent progress in the development of low-inductance, high-current capacitors [40-42] and switches [43-46] opens up big opportunities in the design of new pulsed-power generators, which can directly supply a high-current pulse into a low-inductance *x*-pinch load. For example, PIAF [47-48], built at Cole Polytechnique, France, is a LC generator, based on total of 6, low inductance 180-nF, 50-kV capacitors connected in parallel, which can deliver a 250-kA, 180-ns high-current pulse to an *x*-pinch load without use of any water transmission lines. GenASIS [49], University of California, San Diego, is a 200-kA 150-ns rise time *x*-pinch driver composed of 12 General Atomics, 20-nF, 100-kV capacitors directly arranged around a load section. A compact pulse generator [50], Tomsk, Russia, is based on 4 fast, high-current capacitors and able to supply about 300-kA peak-current with 200-ns rise-time. SPAS [51] current generator at Lebedev Institute, Russia, includes four parallel-connected capacitor-switch assemblies placed in a common tank and delivers about 250-kA peak current with 200-ns rise-time into an *x*-pinch load. MAIZE [52] at University of Michigan, can deliver a huge, up to 1-MA peak-current into *x*-pinch, and even *z*-pinch loads.

Such new pulsed-power generators [47-52] offer many advantages, compared to conventional Marx-based systems. For example, lower capacitor voltage allows one the building of a pulsed-power generator without an isolation oil; being based on low inductance capacitors and switches, such drivers can directly drive a low-inductance *x*-pinch load; no needs for transmission lines allows the elimination of the water conditioning system. Some of these pulsed-power drivers [49, 52] are based on modular, linear transformer driver (LTD) architecture. The basic building block of such systems, called LTD "brick", is usually composed of two capacitors connected in parallel followed by a triggered switch. The brick can be charged up to 100 kV and deliver up to 100-kA peak-current, and it serves as a building block to construct more complicated, pulsed-power systems [53-56]. Most importantly, all these considerations (new low-inductance capacitors and switches, LTD "brick" technology) allows one to design and build compact, non-expansive *x*-pinch drivers which can replace old, conventional, bulky, and expensive pulsed-power systems [22-38].

Our goal is to design and build a compact, portable *x*-pinch radiation source generator with a "good" *x*-ray radiation performance, comparable to standard *x*-pinch drivers. While many experiments have been devoted to studying how *x*-pinch radiation parameters correlate with wire materials, mass, and geometries [2-8, 57-62], they are usually done using one pulsed-power generator, available in each particular case. The literature [2-38, 47-52, 57-63] suggests that a minimum current rise rate, dI/dt, of 1 kA/ns is required to achieve a "good" *x*-pinch *x*-ray radiation performance. For reasonable current rise times, this translates to a current requirement of 150 kA or more. Our initial design uses the recent development of low-inductance, high-current capacitors and switches, as well as the concept of simple, "matched" RLC circuit [64-65] and was described elsewhere [66]. Our final design is based on 2 slow LTD bricks (total of 4 capacitors) combined into a one, solid unit and was reported recently [67]. The new idea behind our *x*-pinch driver design is that it can be truly compact, portable, and available for all activities where an *x*-pinch radiation source is needed. As far as we know, no such portable *x*-pinch drivers currently exist.

## 2. DESIGN AND FABRICATION

### 2.1. Electrical design approach

In our design approach, we are following the description introduced by M. G. Mazarakis & R. B. Spielman [64-65]. For a small generator without long transmission lines, the whole generator with the load can be approximated by a simple RLC circuit. A pulse generator with a "matched" load $R = \sqrt{L/C}$ is, in general, able to produce higher pulse currents with faster rise times compared to the "critically matched" $R = 2\sqrt{L/C}$ case, and is more suitable for the design of a radiographic x-pinch machine [65]. In the "matched" case the peak current and time-to-peak, are given by the following expressions [65]:

$$i_{peak} = 0.546\, V_0/R, \qquad (1)$$

$$t_{peak} = 1.21\sqrt{LC}, \qquad (2)$$

where $V_0$ is the initial voltage of capacitor C.

To prove this approach, we initially designed [66] a compact plasma-radiation-source generator, based on only four, fast high-current capacitors discharged simultaneously in parallel into a "matched" x-pinch load. The simulation shows [66] that the driver supply about 180-kA peak-current with a 150-ns time-to-peak into an x-pinch load.

### 2.2. 2 LTD driver: electrical circuit and simulation

Our final design is based on 2, slow LTD bricks (total of 4 capacitors) that was reported recently [67]. The electrical circuit is shown in Fig. 1 and described below.

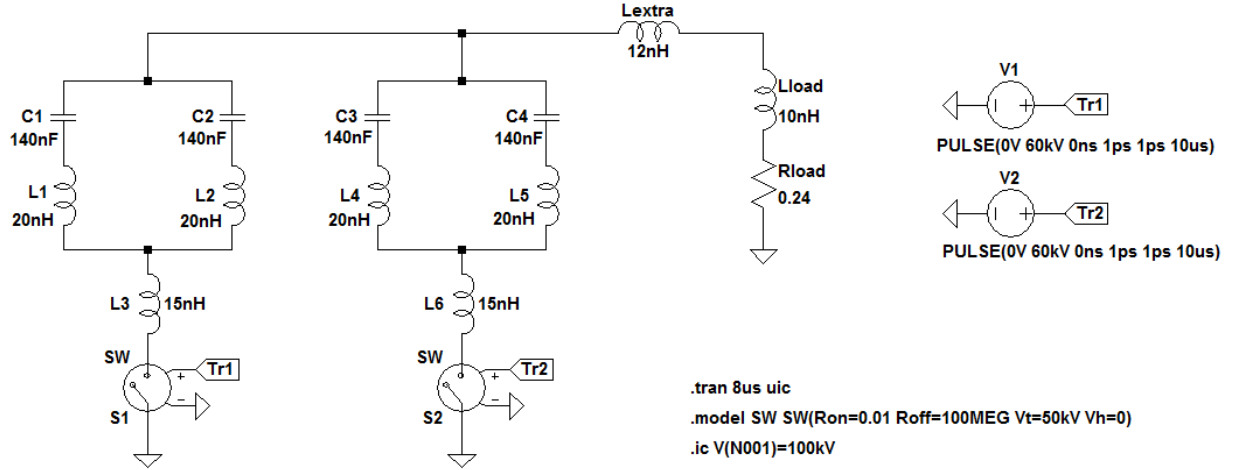

**FIGURE 1.** 2-LTD-brick x-pinch driver electrical diagram.

Each brick is comprised of two GA 35465 capacitors and one multi-gap, air insulated switch. The rated voltage of each capacitor is 100 kV, the rated peak current is 50 kA, capacitance is 140 nF, and inductance is about 20 nH. The inductance of each switch is about 15 nH. The "bare" bricks inductance, $L_{brick}$, is equal to 25 nH and the extra inductance of all elements, $L_{extra}$, which includes extra inductance of bricks placed inside the finite volume, anode-cathode inductance and inductance of all other connections, is estimated to be about 12 nH. The total internal inductance of our x-pinch generator is:

$$L_{driver} = L_{brick}/2 + L_{extra} = (25/2 + 12) \text{ nH} = 24.5 \text{ nH}. \quad (3)$$

The inductance of the *x*-pinch load is estimated to be about 10 nH with a resistance 0.24 Ω. These numbers corresponds to an *x*-pinch load comprised of two 11-mm-long, 40-µm-diameter Mo wires in the initial "cold" state. The real values will be different depending on the electrical parameters of the real "hot" wire state. The total inductance of the whole driver, including the *x*-pinch load, is:

$$L = L_{driver} + L_{x\ pinch} = (24.5 + 10) \text{ nH} = 34.5 \text{ nH}. \quad (4)$$

LTSpice [68] simulations of our driver are presented in Fig 2. The driver can supply 220-kA peak-current with about 170-ns time-to-peak value when it is fully charged to 100 kV. The peak power at the *x*-pinch load is about 11.6 GW, and the energy transferred to the load by the time to peak is about 1.1 kJ. That is about 39% percent of the total 2.8-kJ energy, which can be initially stored inside capacitors. The proposed driver is naturally very efficient and almost half of the initial energy is transferred to the load by the time to first peak.

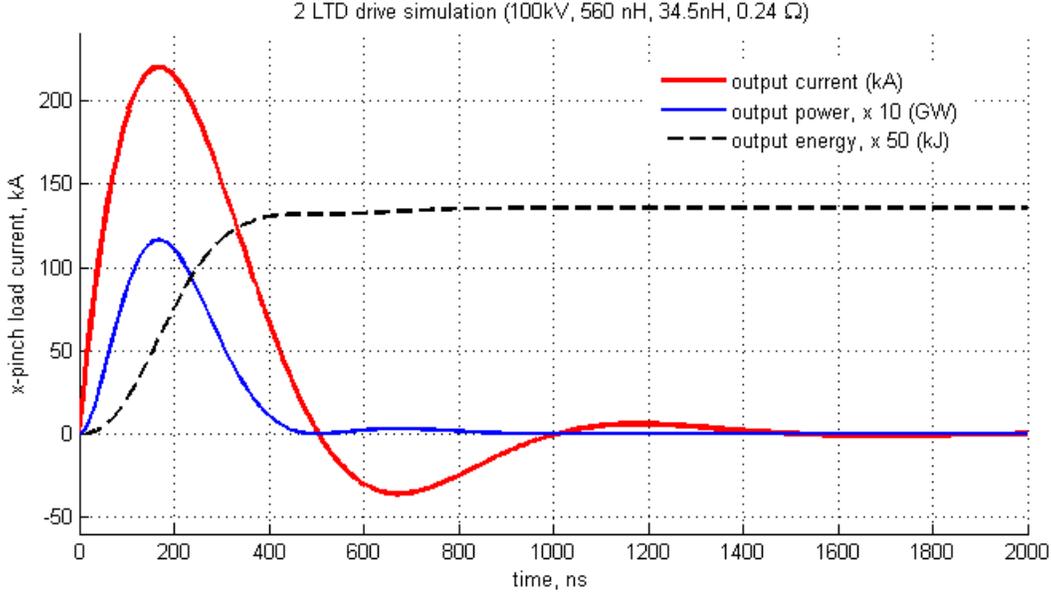

**FIGURE 2.** LTSpice simulation of 2-LTD-brick driver. Red line is the output load current, blue is the output power, black (dotted) is energy transferred to the load.

The parameters simulated above of the *x*-pinch generator, such as a peak current and a rise time, can be easily predicted and verified by the simple RLC model described above. The "matched" *x*-pinch load value of our generators equals 0.25 Ω and, according to formulas (1, 3), the peak current and rise time can be calculated as:

$$i_{peak} = 0.546 \times 100 \text{ kV}/0.25\ \Omega = 218 \text{ kA}, \quad (5)$$

$$t_{peak} = 1.21\sqrt{34.5 \text{ nH} \times 560 \text{ nF}} = 168 \text{ ns}, \quad (6)$$

which are in a good agreement with the values simulated above.

## 2.3. 2 LTD driver: mechanical design and fabrication

The mechanical design of our *x*-pinch *x*-ray generator is based on the desire to make the driver compact and portable and to minimize the total driver inductance. We used 2, slow NRL LTD bricks [69] combined into one, solid unit. The *x*-pinch driver assembly is pictured in Fig. 4. Bricks are placed inside the steel housing with ground and output plates on opposite sides. The vacuum chamber is placed at the top of the output plate and is designed to withhold a low vacuum needed for *x*-pinch *x*-ray generation. It is composed of a 6" in diameter, cylindrical body and two 2-3/4" output ports.

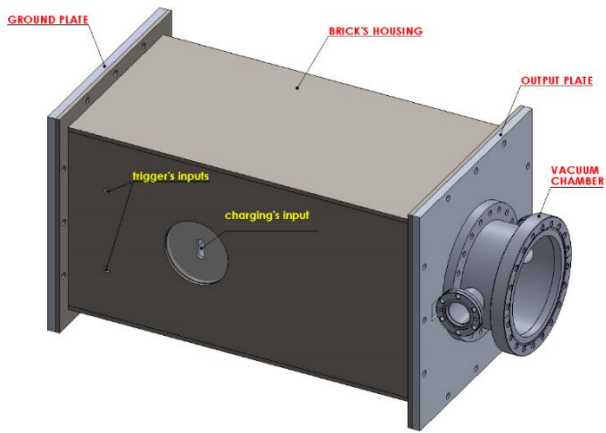 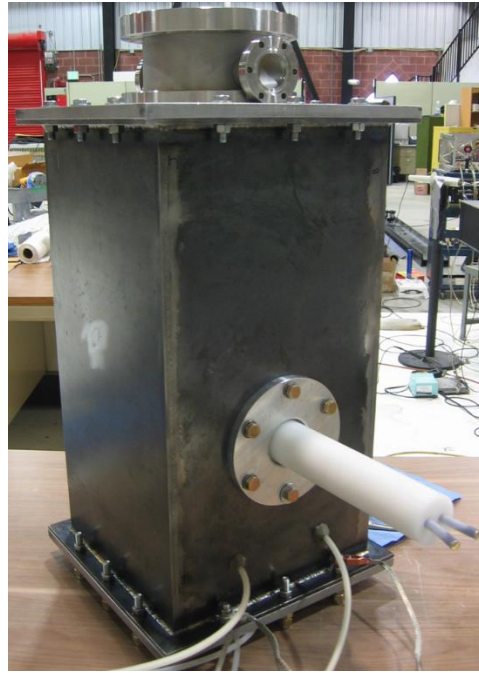

**FIGURE** 3. 2-LTD-brick *x*-pinch *x*-ray generator, general view. Left - drawing, right – image.

The output section of *x*-pinch *x*-ray generator is presented in Fig. 4. The high voltage plate combines 2 bricks and feeds the output current into the load section. A ¾" isolating acrylic plate is placed between the high voltage and output plates. A 2" long, 2" diameter cathode section is attached to the high voltage plate. The anode section is 1-3/4" long, has the 2-1/2" inner diameter and has two output *x*-ray windows aligned with vacuum chamber output ports.

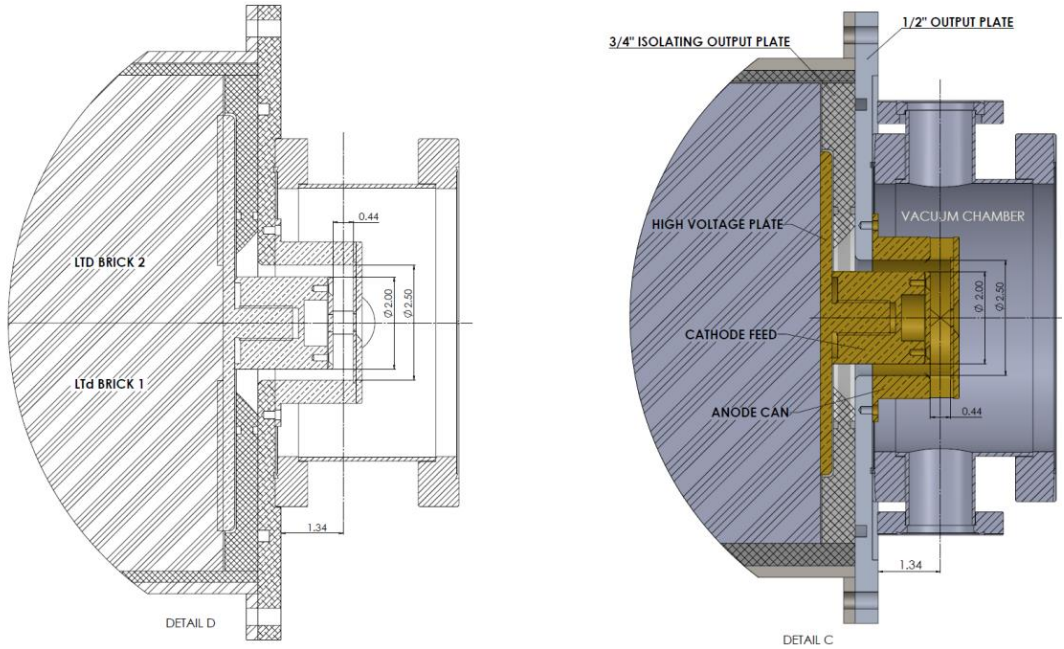
**FIGURE 3.** The output cross-section view of 2 LTD bricks *x*-pinch *x*-ray generator.

To minimize the driver inductance, the anode-cathode gap was reduced to 6.3 mm. With such a small A-K gap, the power feed must operate in the self magnetically insulated mode. The $v \times B$ Lorentz force induced by a current pulse will turn back electrons emitted from the cathode and stop the voltage breakdown in the inter-electrode space. The minimum operating current, according to [47], can be estimated to be:

$$I_{\min} = 0.64 \frac{\sqrt{V(\text{kV})}}{\ln(R_a / R_c)}. \tag{7}$$

where $R_a$ and $R_c$ are anode/cathode diameters. With the present geometry, the minimum operating current becomes:

$$I_{\min} = 2.87 \sqrt{V(\text{kV})}. \tag{8}$$

At a maximum anode-cathode voltage of 55 kV, for the magnetic self-insulating principle to work, the minimum operating current becomes 21 kA. This is well below the predicted (5) 220-kA value.

All *x*-pinch driver parts were carefully designed using SolidWorks 2013 [70], 3D solid modeling engineering software. A total of 41 drawings needed for *x*-pinch driver fabrication were generated, and are stored at IAC backup server and can be requested, if needed.

# 3. DRIVER TESTING AND CHARACTERIZATION

## 3.1. Pulsed-current diagnostic

An electrostatically shielded Rogowski coil [71-72] with a very good signal-to-noise ratio was designed and placed at x-pinch driver output section to measure the total current delivered into an x-pinch load. Our Rogowski coil has a total of 29 turns with 1-inch winding spacing. To calibrate the Rogowski coil, the set-up (shown in Fig. 5) was used. A total of 5 fast, 30 nF, HV capacitors were initially charged to 5 kV to produce high-current oscillations. A 0.1003-$\Omega$ calibration current viewing resistor (CVR) was placed at the end of the metal strip line. This set-up allowed us to reach a 360-ns time-to-first-peak value which is close to the *x*-pinch driver parameter.

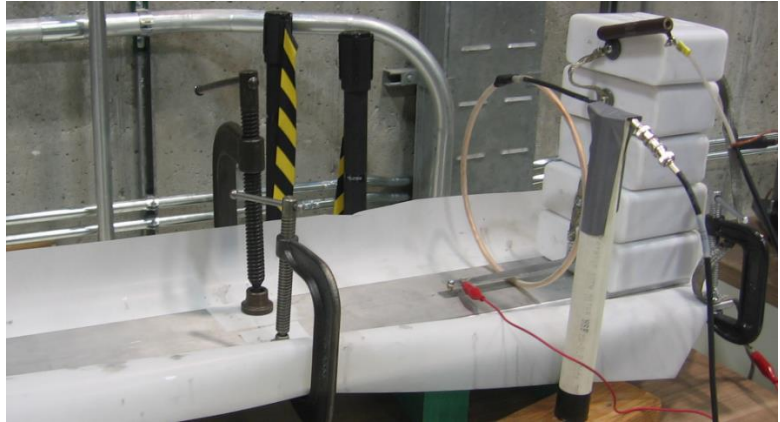

**FIGURE 4.** Rogowski coil calibration set-up.

The calibration data are shown in Fig. 6. The output Rogowskii coil signal is proportional to dI/dt and should be integrated to get a current waveshape. The special calibration method was developed to reasonably fit the Rogowski integral to CVR current. The Rogowski coil calibration factor is found to be 2.51.

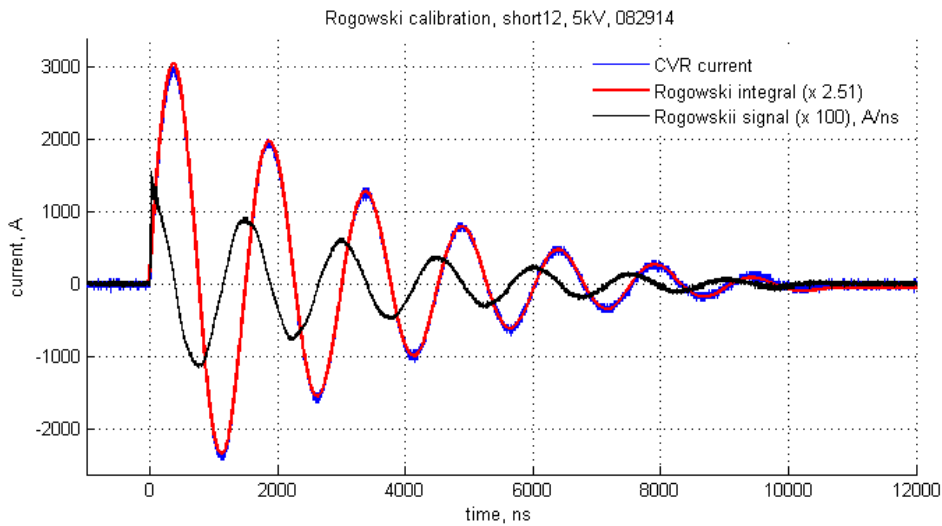

**FIGURE 5.** Rogowski coil calibration data; blue line – CVR current, black line – Rogowski output signal in V; red line – Rogowski integral normalized to match CVR current.

## 3.2. Short circuit test data

Several configurations of a 2-LTD-brick *x*-pinch driver were tested before a good "working" configuration was found. To show the general progress as the driver configuration was improved from test to test, test data are briefly summarized in Table 1. Initial tests (test 1-3) were done with 1-Ω resistor immerged in dielectric oil. Tests 4 and 5 were performed with a salted water load filled the anode-cathode gap and the lower part of the cathode feed. To reduce the total driver inductance, starting from test 5, the Cu sheet was placed between two bricks inside the brick housing.

**TABLE 1.** 2-LTD-brick *x*-pinch driver tests data.

|        | Date, 2014 | Load           |    | Short # /kV | $I_{peak}$ (kA) | $T_{peak}$ (ns) | $dI/dt_{max}$ (kA/ns) | Comments and common problems |
|--------|------------|----------------|----|-------------|-----------------|-----------------|-----------------------|------------------------------|
| Test 1 | May 12-13  |                |    | 3/76        | 47              | 300             | 0.40                  | Common charging line, pre-fire, HV plate spark, only 1 brick is fired; |
| Test 2 | Jun 11     | 1 Ω HVR        |    | 1/80        | 72              | 420             | 0.35                  | Separate charging line, non-pre-fire, HVR spark, 2 bricks are fired differently in time; |
| Test 3 | July 17    |                |    | 2/78        | 40              | 290             | 0.30                  | Same as test 3, but no HVR spark, 2 bricks separation time ~ 1.1 us; resister is expanded; |
| Test 4 | July 24    | Salted water   | R1 | 6/80        | 153             | 382             | 0.71                  | Sparking at the end of trigger line eliminated, all bricks are fired simultaneously; |
|        |            |                | R2 | 11/80       | 150             | 386             | 0.66                  |                              |
|        |            |                | R3 | 18/80       | 144             | 382             | 0.60                  |                              |
|        |            |                | R4 | 22/80       | 124             | 377             | 0.58                  |                              |
|        |            |                | R5 | 29/80       | 115             | 365             | 0.56                  |                              |
|        |            |                | R6 | 39/80       | 100             | 360             | 0.53                  |                              |
| Test 5 | Aug 1      |                | R  | 4/80        | 170             | 335             | 0.91                  | Cu sheet is placed between two bricks to reduce the system inductance, 2 PT-55; |
| Test 6 | Aug 14     | Wire load in oil |  | 6/80        | 185             | 370             | 0.86                  | Configuration is the same as in test 5; |

The final test 6, for short event 3, is discussed in more details below. The bricks were initially charged to 80 kV and the load was a 1.48-cm-long, 2.9-mm-diameter Ni-wire. The experimental data are presented in Fig. 7. The Rogowski peak current is about 185 kA with 220-ns, 10-90%, rise time. The corresponding current rise rate is 0.84 kA/ns.

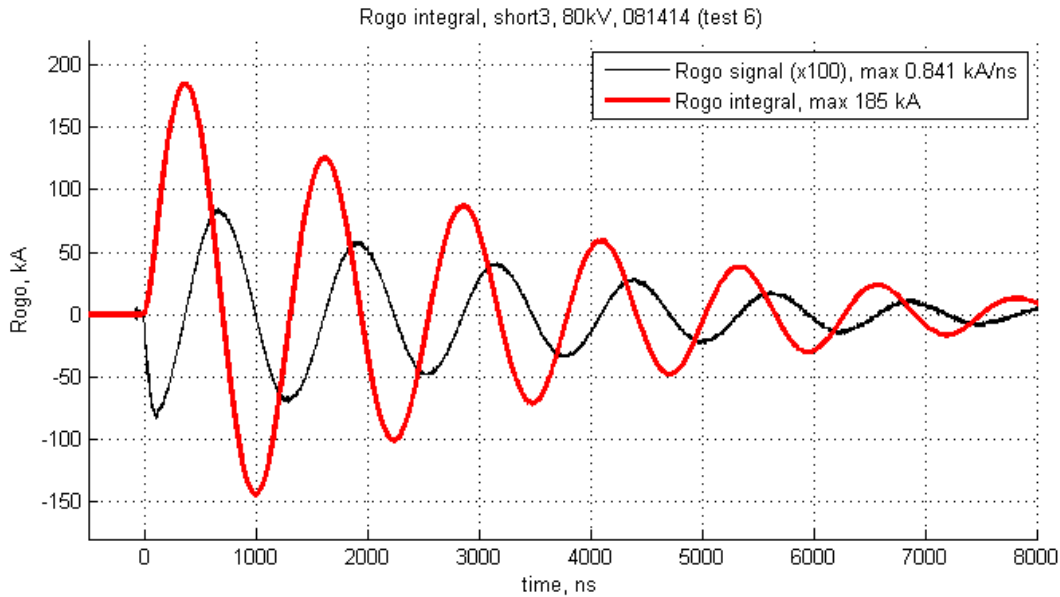

**FIGURE 6.** 2-LTD-brick driver test data with wire load, 08/14/2014, short 3, 80 kV; black line – Rogowski output current, kA/ns, red line – Rogowski current integral, kA.

To better understand the driver parameters, LTSpice simulations were compared with experimental data and results are presented in Fig. 8. The total driver inductance and resistance values were varied until good agreement with measured oscillation time and decay constant was achieved. The inductance and resistance values were found to be 69 nH and 0.042 Ω, correspondingly. However, simulations cannot reproduce the initial capacitors charge 80 kV. Most likely, there were some losses inside the driver which effectively reduced the initial voltage down to about 71 kV. After 5 µs, the experimental current starts to deviate from simulated one. This can be explained by changes in the switch resistance, which starts to increase at the end of the current pulse.

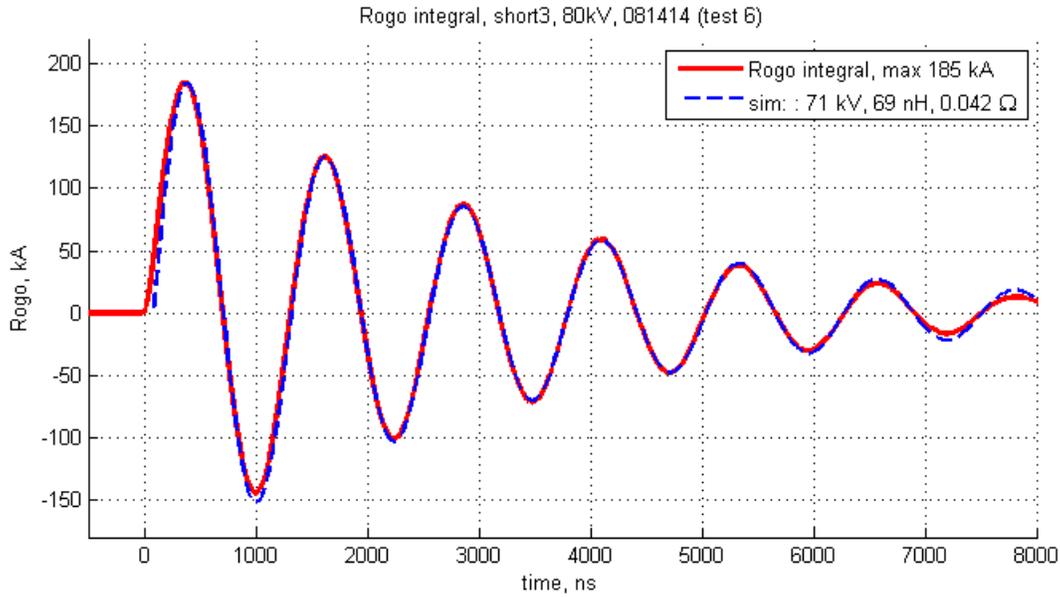

**FIGURE 7.** Comparison of experimental current with LTSpice simulation; red line – Rogowski current, 08/14/2014, short 3, 80 kV; blue (dotted) lines – LTSpice simulation.

The total internal inductance of our 2-LTD-brick x-pinch driver can be estimated as follow. The inductance value found above, 69 nH, is the total driver inductance, which includes the wire load value. The wire load inductance can be estimated as:

$$L_{load} = 2h \ln \frac{r_0}{r_1} = 9.12 \text{ nH} . \qquad (9)$$

where, h is the wire length, 1.48 cm, $r_1$ is the wire diameter, 1.45 mm, and $r_0$ is the inner anode diameter, 31.75 mm. So, the total internal driver inductance should be equal to:

$$L_{driver} = L_{measured} - L_{load} = (69 - 9.12) \text{ nH} = 60 \text{ nH}. \qquad (10)$$

### 3.3. Revised driver model and projected x-ray performance

The 2-LTD-brick driver model was revised based on the short-circuit test data. The electrical circuit diagram, presented in Fig 1, was modified to match the total internal driver inductance to the measured 60-nH value. The total driver inductance, including *x*-pinch load, was 70 nH. The simulations are presented in Fig. 9 and briefly discussed below.

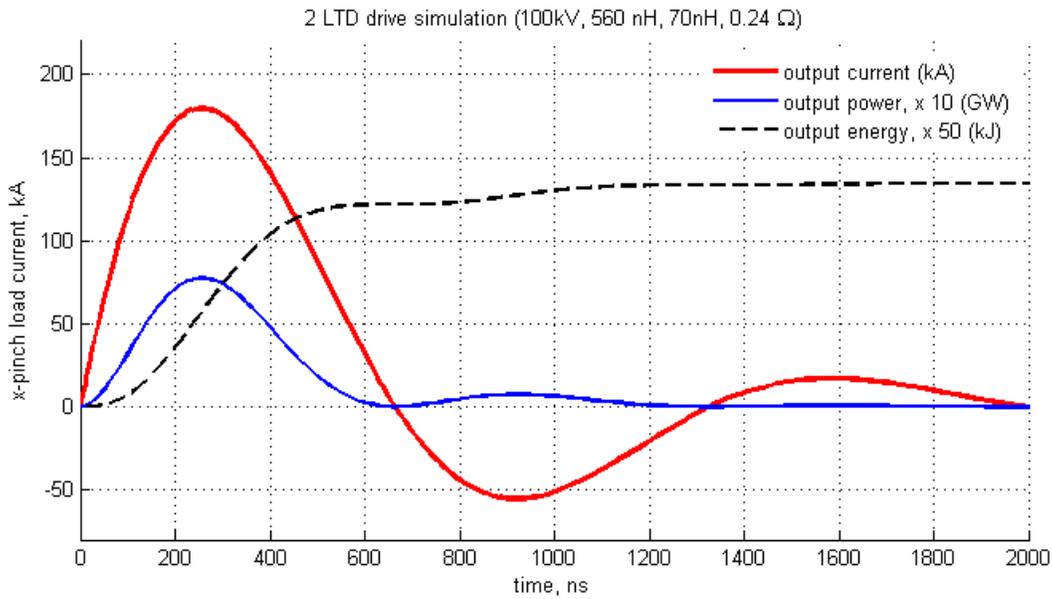

**FIGURE 8.** LTSpice simulation of 2 LTD bricks driver. Red line is the output load current, blue is the output power, black (dotted) is energy transferred to the load.

As can be seen, the driver, when it is fully charged to 100 kV, can supply 180-kA peak-current with about a 150-ns, 10-90%, rise time. The corresponding current rise rate, dI/dt, equals 1.2 kA/ns, which is close to the measured value of 0.84 kA/ns when it scaled to 100 kV. The peak power at the x-pinch load is about 7.7-GW, and the energy transferred to the load by the time to peak is about 1.2 kJ. That is 43% percent of the total 2.8 kJ of energy initially stored inside capacitors.

The projected *x*-pinch performance of our 2 LTD *x*-pinch driver is presented in the Fig. 9. The data are taken from PIAF generator [47] which can supply about 250-kA peak current to a 6-nH inductive load with a 180-ns current rise time. For the load composed of two 25-μm Mo wires, the projected *x*-ray performance is at least 250-500 mJ in a pulse 1.2-2.0-ns FWHM for photon energy above 700 eV [47].

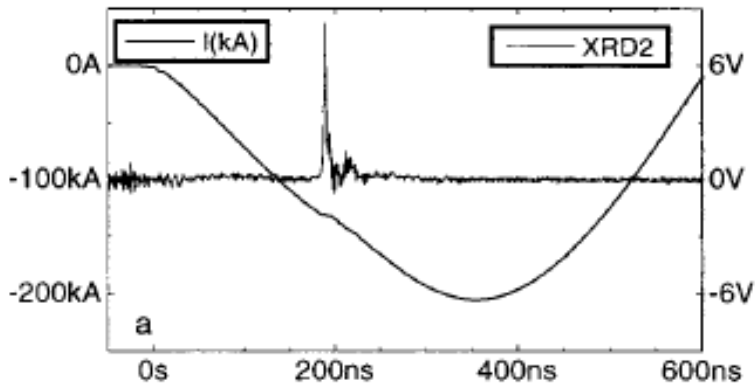

**FIGURE 9.** Projected *x*-pinch performance from PIAF pulsed power generator. The load is two 25-μm Mo wires and *x*-ray data is with a 3-μm Mylar filter.

# 4. SUMMARY AND FUTURE WORKS

The objective of this proposal is to develop a new, compact and truly portable *x*-pinch radiation source generator with a "good" *x*-ray radiation performance; the radiation source should be fast, small, and bright; and should be comparable to other standard *x*-pinch drivers. Below is a short summary of what has been done and what is expected to make this research project a success:

- Literature suggests that there is a need for compact and portable *x*-pinch radiation sources for use in a variety of applications in physics, biology and industry. The current rise rate, dI/dt, of 1 kA/ns or more is required for a "good" *x*-pinch radiation performance.

- A new compact and portable *x*-pinch driver was recently designed and constructed at IAC. Our final design is based on 2 "slow" LTD bricks combined into one, solid unit. The driver inductance was minimized and the output load current was maximized. The size of our driver is only 28x13x14 inches and it weighs about 200 pounds.

- The test data reveals that the driver can supply 185-kA peak-current into a short wire load. The rise time, 10-90%, is about 220-ns and the corresponding current rise rate is 0.84 kA/ns. The measured total internal inductance of our driver is about 60 nH.

- The simulation, based on the revised driver model, shows that the driver can deliver about 180-kA peak current into an *x*-pinch load with a 150-ns, 10-90%, rise time. The corresponding current rise rate is 1.2 kA/ns.

- *X*-ray radiation parameters of constructed 2-LTD-brick *x*-pinch driver will be measured and characterized for different wire materials and geometries. The data will be compared with other well-known *x*-pinch drivers with a similar current-rise-rate.

- The performance of constructed *x*-pinch driver as an *x*-ray backlighting tool will be evaluated and reported. The study of an exploding wire is among all possible applications of our constructed 2-LTD-driver. This study is currently under progress at the IAC, and will help to better understand the EOS of a warm dense matter. We will image this plasma with our *x*-pinch and will characterize the spatial and temporal resolution of our radiation source.

- The low cost of operation, compactness, and portability of the constructed 2 LTD *x*-pinch driver will make it available for many experiments where a bright, fast *x*-ray source is needed. As far as we know, no such portable *x*-pinch drivers currently exist.